\documentclass[largeformat]{interact}

\usepackage{epstopdf}% To incorporate .eps illustrations using PDFLaTeX, etc.
\usepackage[caption=false]{subfig}% Support for small, `sub' figures and tables
\usepackage[doublespacing]{setspace}% To produce a `double spaced' document if required
\usepackage{graphicx}% Include figure files
\usepackage{dcolumn}% Align table columns on decimal point
\usepackage{bm}% bold math
\usepackage{amsmath}
\usepackage{amssymb}
\usepackage{latexsym}
\usepackage{epsfig}
\usepackage{amsbsy}
\usepackage{array}
\usepackage{amssymb}
\usepackage{setspace}
\usepackage{bm}
\usepackage{epstopdf}
\usepackage{siunitx}
\usepackage{url}
\usepackage[font=normalsize,labelfont=bf]{caption}

\usepackage{xcolor}
\usepackage[numbers,sort&compress]{natbib}% Citation support using natbib.sty
\makeatletter% @ becomes a letter
\def\NAT@def@citea{\def\@citea{\NAT@separator}}% Suppress spaces between citations using natbib.sty
\makeatother% @ becomes a symbol again

\theoremstyle{plain}% Theorem-like structures provided by amsthm.sty

\theoremstyle{definition}

\theoremstyle{remark}

\begin{document}

\articletype{Original reports}% Specify the article type or omit as appropriate

\title{Pressure effect on the order-disorder transformation in L1$_0$ FeNi}

%\name{Li-Yun Tian Author\textsuperscript{a,b}\thanks{CONTACT A.~N. Author. Email: latex.helpdesk@tandf.co.uk}, Olle Eriksson\textsuperscript{b,c} and Levente Vitos\textsuperscript{a,b,d}}
\author{
\name{Li-Yun Tian\textsuperscript{a,b}\thanks{CONTACT Li-Yun Tian. Email: liyunt@kth.se}, Olle Eriksson\textsuperscript{b,c} and Levente Vitos\textsuperscript{a,b,d}}\thanks{CONTACT Levente Vitos. Email: levente@kth.se}
\affil{\textsuperscript{a}Applied Materials Physics, Department of Materials Science and Engineering, Royal Institute of Technology, Stockholm SE-100 44, Sweden; \textsuperscript{b}Department of Physics and Astronomy, Division of Materials Theory, Uppsala University, Box 516, Uppsala SE-751 20, Sweden;
\textsuperscript{c}School of Science and Engineering, \"{O}rebro University, \"{O}rebro, Sweden;
\textsuperscript{d}Research Institute for Solid State Physics and Optics, Wigner Research Center for Physics, Budapest H-1525, Hungary}
}

\maketitle

\begin{abstract}
%The ordered phase of the FeNi system is known to have promising magnetic properties as a rare-earth free permanent magnet. Understanding the parameter space that controls the order-disorder transformation is important in order to find growth conditions that stabilize the $L1_0$ phase of FeNi. In this work, the magnetic properties and chemical order-disorder transformation of FeNi are investigated as a function of lattice expansion using first-principles alloy theory. The largest volume expansion considered here is 29 \% which corresponds to a pressure of \SI{-25}{\giga\pascal}. The thermodynamic and magnetic calculations are performed in terms of a long-range order parameter, which is then used to find the ordering temperature as a function of pressure. We show that negative pressure promotes ordering and thus synthetic routes that involve an increase of the volume of FeNi is expected to expand the stability field of the $L1_0$ phase.\\

The ordered phase of the FeNi system is known to have promising magnetic properties as a rare-earth-free permanent magnet. Understanding the parameter space that controls the order-disorder transformation is important to find growth conditions that stabilize the $L1_0$ phase. Magnetic properties and chemical order-disorder transformation of FeNi are investigated as a function of lattice expansion using first-principles theory. Thermodynamic and magnetic calculations are performed using long-range order parameters, which is used to find the ordering temperature. Negative pressure promotes ordering and thus synthetic routes involving an increase in volume is expected to expand the stability field of the $L1_0$ phase.\\

\noindent\textbf{IMPACT STATEMENT}

\noindent The order-disorder transition temperature of tetragonal FeNi alloy increases with (negative) pressure, which is expected to promote the stability of tetrataenite as the primary candicate for rare-earth-free permanent magnets.

\end{abstract}


%\section{Introduction}

Tetragonal FeNi alloy as a candidate permanent magnet material, exhibiting high uniaxial magnetic anisotropy and large magnetic moment, has generated increasing interest \cite{Kojima2014, Skomski2013Future, Lewis2014Inspired}, especially since the discovery of single phase tetrataenite (NWA 6259) in 2010. Tetrataenite is a rare-earth-free and low-cost permanent magnet which is expected to be applied widespread in future electromagnetic devices \cite{Lewis2014magnete, Yodoshi2018effects}. However, there is the serious problem of the order-disorder transition temperature (320$^o$C) of tetragonal FeNi, which prevents efficient synthesis routes of this phase, due to the slow atomic diffusion at such a low temperature. Designed to overcome this deficiency, specific tailored synthesis techniques are extremely promising. Techniques that in general increase the order-disorder temperature include the promotion of atomic diffusion, via e.g. atomic vacancies or defects, induced distorted lattice \cite{Frisk2017strain}, and third element effect \cite{Kojima2014Co}. 
Modern monatomic layer deposition technique has been used to fabricate $L1_0$ FeNi thin films on a substrate with a small structural match \cite{Shima2007film, Kojima2014film}. Unfortunately, this structural realization has decreased long-range order parameter and magnetic anisotropy energy, $K_u$.
Recently, Goto et al. synthesized $L1_0$ FeNi with a high degree of order, of about 0.71 through nitrogen insertion and topotactic extraction (NITE) technology which is the highest degree of order reported so far \cite{Goto2017}. 

As soon as practical synthesis and processing techniques are developed to access the tetrataenite phase in bulk amounts and extend its stability field beyond room temperature, the excellent potential of $L1_0$ FeNi can be realized for permanent magnet applications.
Until now, the studies on how to stabilize tetragonal FeNi have been inconclusive. Specifically, there have been few reports that clarify the dependence of the physical properties on volume contraction or expansion. However, it has been confirmed that the chemical and magnetic ordering transitions are interdependent of each other \cite{Jena2014study}. In this work we apply a Monte Carlo (MC) method to treat the spin coupled Hamiltonian of the magnetic ordering effects and density functional theory (DFT) to treat the chemical ordering effects. 

In our previous work, we have accurately demonstrated an order-disorder transition in  $L1_0$ FeNi alloy evaluated from the local atomic rearrangements for chemical ordering \cite{Tian2019FeNi}. Our approach represents an efficient way to describe the ordering transformation based on \emph{ab initio} theory. 
In the present work, we report the chemical order-disorder transformation and magnetic transition of tetragonal FeNi alloys under various negative pressures. Our goal is to investigate how chemical order-disorder and magnetic transitions change as a function of (negative) pressure, as well as the correlations between magnetic properties and microstructure.
The primary conclusion of our investigation is that negative pressure (e.g. chemical volume expansion) is a potential way to increase the order-disorder transition temperature of $L1_0$ FeNi. 

In order to check the negative pressure behavior, we have undertaken first-priciples calculations employing the generalized gradient approximation (GGA) within the exact-muffin-tin orbitals (EMTO) method \cite{Vitos2000a, Vitos2001c, Vitos2007book} to evaluate electronic and magnetic properties of tetragonal FeNi alloy.
First-principles Gibbs free energy calculations were used to describe the thermodynamic properties of FeNi alloys as a function of a long-range order parameter $\eta$.
The ground-state properties and vibrational and electronic contributions to the Gibbs free energy were computed within the EMTO and magnetic free energy was simulated using the MC approach, as implemented in the Uppsala atomistic spin dynamics (UppASD) program \cite{Skubic2008MC}. The exchange parameters needed in the MC simulations were calculated using EMTO. Details regarding the computational scheme are the same as those used in our previous work \cite{Tian2019FeNi}. 

%\subsection{The \textsf{Interact} class file}\label{class}

%The \texttt{interact} class file preserves the standard \LaTeXe\ interface such that any document that can be produced using \texttt{article.cls} can also be produced with minimal alteration using the \texttt{interact} class file as described in this document.

%\subsection{Submission of manuscripts prepared using \emph{\LaTeX}}

%\section{Using the \texttt{interact} class file}

The order-disorder transformation was studied with body-centered tetragonal (bct) crystal structure with two sublattices corresponding to $i$ and $j$ sites. Using the single-site coherent-potential approximation (CPA) \cite{Soven1967coherent, Gyorffy1972coherent} to describe the random structures, both $i$ and $j$ sites were occupied by two chemical types of atom. In the process of expanding the volumes of systems with various degrees of chemical order, the tetragonal distortions ($c/a$) were fixed at the corresponding long-range order structures calculated in our previous work \cite{Tian2019FeNi}.

The Gibbs free energy is considered as a function of temperature \emph{T}, and pressure \emph{P} at different long-range order parameter $\eta$ values:
\begin{equation}
\begin{split}
G (T, P, \eta) =  E_{0K} (\eta) - T S_{conf} (\eta) + F_{vib} (P, T, \eta) \\
+ F_{el} (P, T, \eta) + F_{mag} (P, T, \eta) + PV,
\end{split} \label{eq:freeE}
\end{equation}
where \emph{V} is the equilibrium volume corresponding to pressure \emph{P}, $E_{0 K}$ is the internal energy per unit cell at 0 K, $S_{conf} (\eta) = - \frac{k_B}{N} \large [ 2 \times ( c + \frac{1}{2} \eta) \times \ln (c + \frac{1}{2} \eta) + 2 \times (c - \frac{1}{2} \eta) \times \ln (2 - \frac{1}{2} \eta) + 2 \times (1 - c - \frac{1}{2} \eta) \times \ln (1 - c - \frac{1}{2}) + 2 \times (1 - c + \frac{1}{2} \eta) \times \ln (1 -c + \frac{1}{2} \eta) \large ]$ is the configurational entropy, and $F_{vib}$, $F_{el}$ and $F_{mag}$ are the vibrational, electronic and magnetic free energies, respectively. The vibrational free energy $F_{vib} (P, T, \eta)$ was described by Debye model with the Debye temperatures determined from the tetragonal elastic parameters calculated for the ferromagnetic state. The electronic contribution to the free energy was determined by the electronic density of states at the Fermi level.
The magnetic contribution to the Gibbs free energy was obtained as $F_{mag} \approx - T S_{mag} \approx - T \frac{\partial \langle H_{mag} \rangle}{\partial T}$ where the Heisenberg Hamiltonian was described as $H_{mag}$ = $ - \sum_{i \neq j} J_{ij} \mu_{i} \mu_{j} \widehat{e}_i \widehat{e}_j$ where the exchange interaction $J_{ij}$ and the local magnetic moments $\mu_{i}$ and $\mu_{j}$ on sites $i$ and $j$ were obtained from EMTO calculations.

\begin{figure}[ht]
	\centering
	\includegraphics[width=1.0\linewidth]{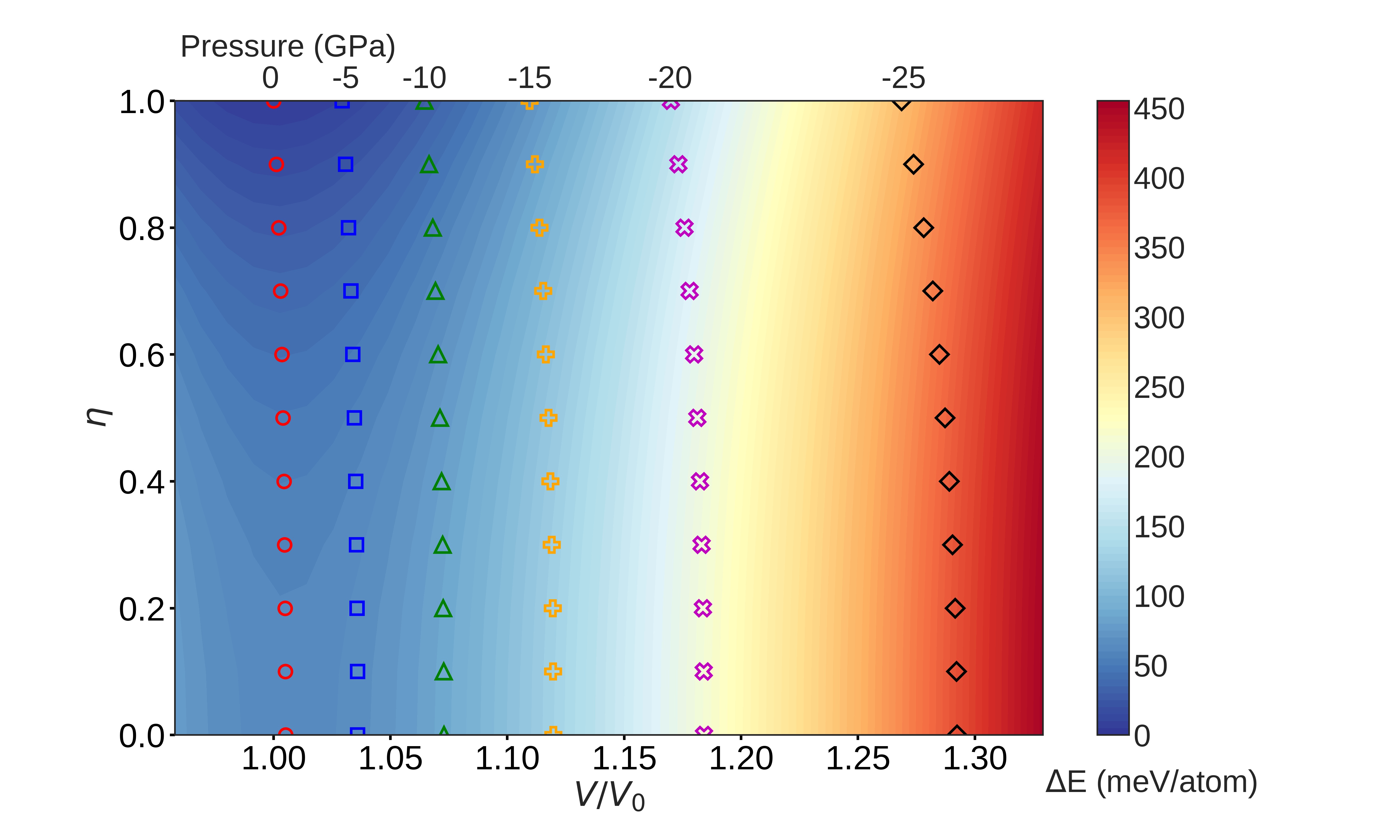}
	\caption{(Color online) The \emph{$V/V_0$} ratios of tetragonal FeNi alloys corresponding to constant pressures (shown at the top of the figure) as a function of ordering parameter. $V_0$ represents the equilibrium volume of fully ordered FeNi at \SI{0}{\giga\pascal}. The colored symbols show the equilibrium volumes at constant pressure (0, -5, -10, -15, -20 and \SI{-25}{\giga\pascal}) as a function of ordering parameter. The color gradient shows the changes of internal energies relative to the \SI{0}{\giga\pascal} total energy of $L1_0$ FeNi.}\label{fig:presure_volume}
\end{figure}

At \SI{0}{\giga\pascal}, the equilibrium volumes and the tetragonal distortions ($c/a$) of FeNi alloys were obtained by mapping the minimum of the total ground-state energies as a function of $\eta$. The pressure-dependent equilibrium volumes were obtained from the free energies at the corresponding $c/a$ ratios at \SI{0}{\giga\pascal}. Once the corresponding $c/a$ ratios were fixed, the equilibrium volumes of fully and partially ordered FeNi alloys were fitted by the Birch-Murnaghan equation of state \cite{Murnaghan1944,Birch1947}. For the external pressure we considered values between 0 and \SI{-25}{\giga\pascal} with a step of \SI{5}{\giga\pascal}. We note that the largest negative pressure considered here corresponds approximately to 29 \% lattice expansion caused by 33.3 at.\% interstitial N \cite{Goto2017}.

In Fig. \ref{fig:presure_volume}, the equilibrium volumes of ferromagnetic FeNi alloys are shown at different ordering states and negative pressures (as colored symbols). The color gradient shows the difference of the internal energies at different site occupations relative to the \SI{0}{\giga\pascal} total energy of fully ordered FeNi configuration. It should be observed that for each pressure, the ferromagnetic state is stabilized with increasing long-range order parameter. In addition, the site occupation induced change in the volume at constant pressure is relatively small. That is because the atomic sizes of Ni and Fe are very similar.

\begin{figure}[ht]
	\centering
	\includegraphics[width=0.84\linewidth]{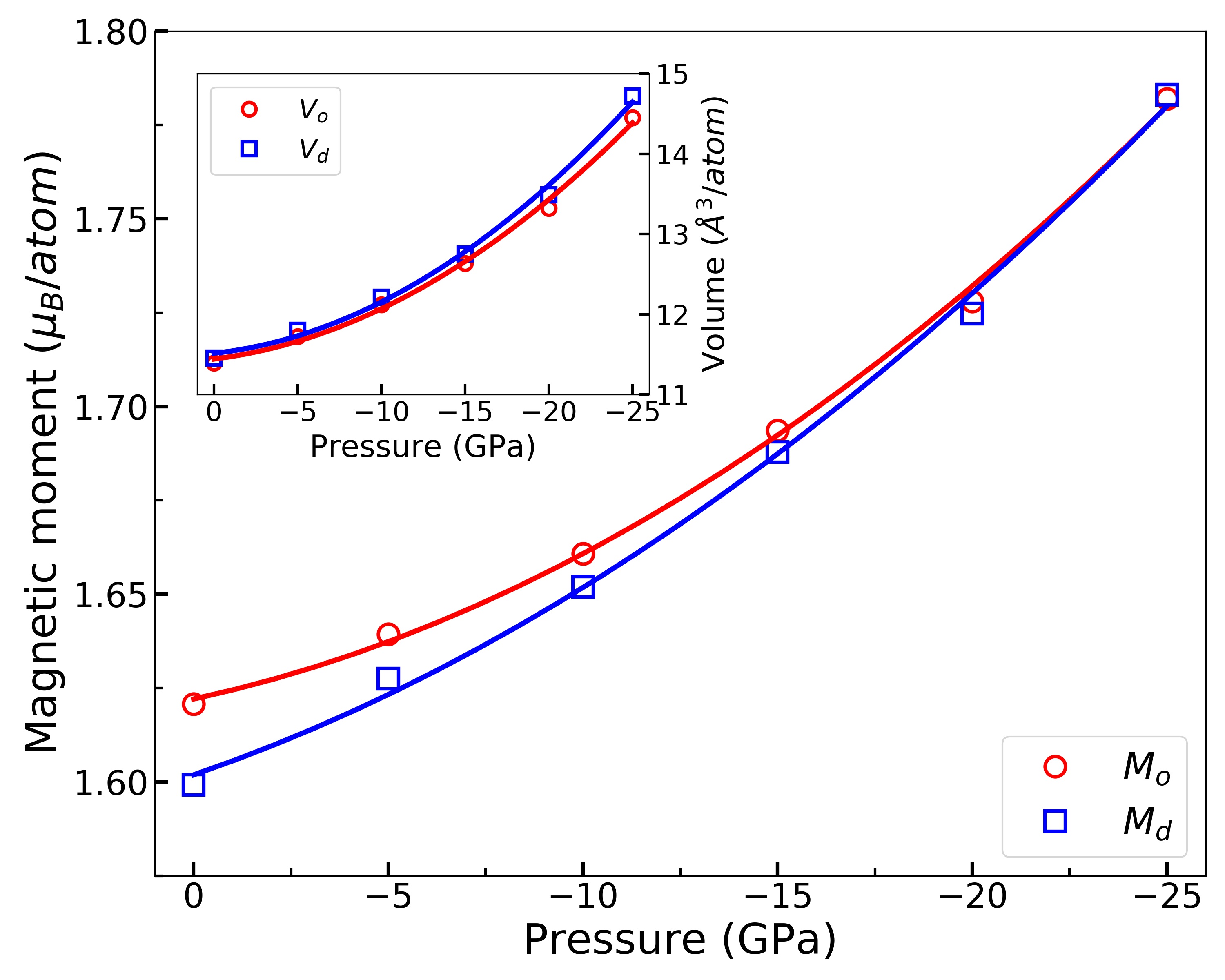}
	\caption{(Color online) Pressure dependence of the average magnetic moments and volumes (inset) for the fully ordered and disordered FeNi alloys. }\label{fig:fig3_mag}
\end{figure}

The spin magnetic moments of fully ordered and random FeNi are shown as a function of pressure in Fig. \ref{fig:fig3_mag}. The magnetic moments are calculated by averaging the magnetic moments of Fe and Ni sites in the bct structure. It is well-known that itinerant ferromagnetism is caused by unequal occupancy of the majority-spin and minority-spin in the density of states (DOS). Expanding the atomic volume causes transition between the majority-spin and minority-spin which yields increased magnetic moments. The average magnetic moment of fully ordered FeNi is 1.621 $\mu_B$ and 1.782 $\mu_B$ per atom at \SI{0} and \SI{-25}{\giga\pascal}, respectively, while 1.599 $\mu_B$ and 1.783 $\mu_B$ per atom for the random phases. The difference between the magnetic moments of tetragonal and random structures becomes smaller with increasing (negative) pressure due to the fact that the volume of random phase is more sensitive to pressure than that of the ordered phase, as shown in the inset of Fig. \ref{fig:fig3_mag}. However, at the same pressures, the magnetic moments of the ordered phases with smaller volumes are slightly larger than those of corresponding random structures with bigger volumes. This is due to the increased minority-spin in DOS of $L1_0$ FeNi as compared to the random phase \cite{Tian2019FeNi}.

\begin{figure*}[ht]
	\centering
	\includegraphics[width=1.0\linewidth]{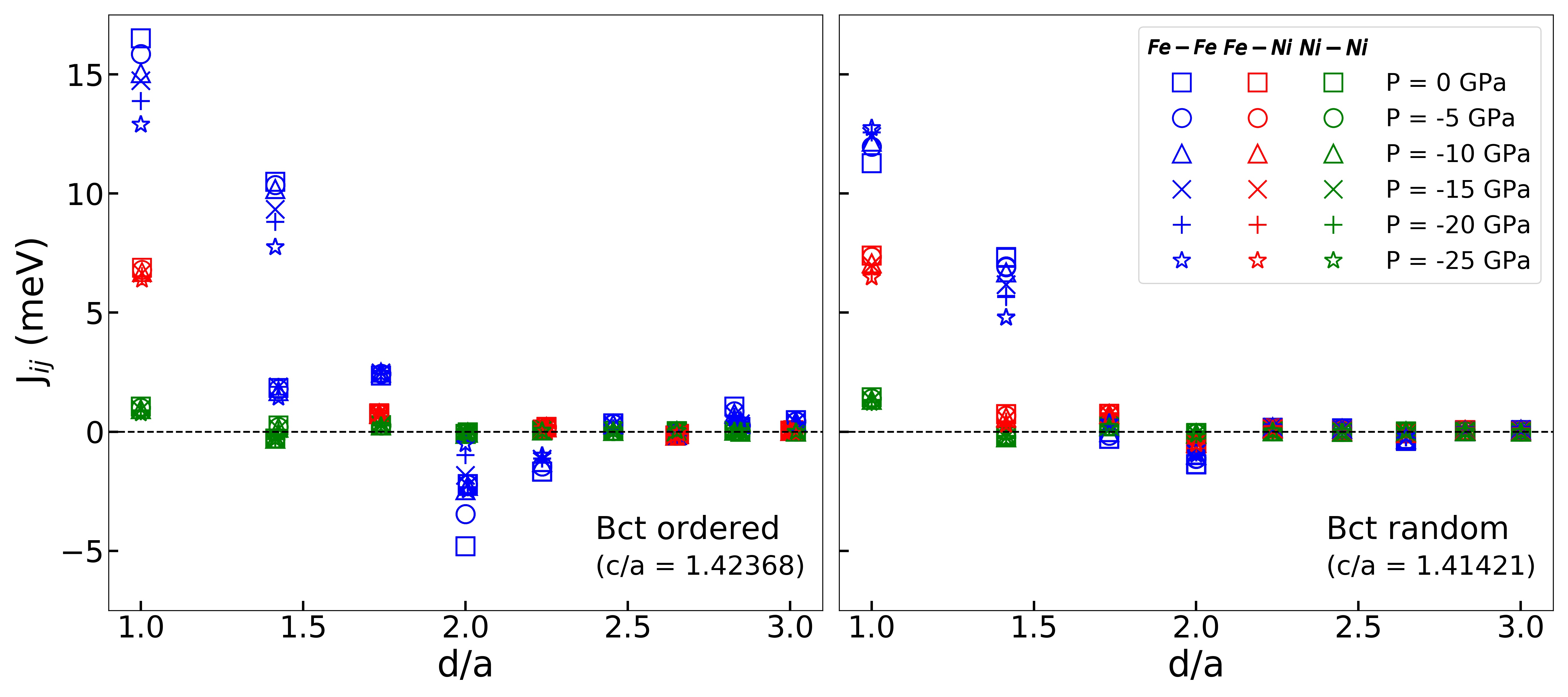}
	\caption{(Color online) Magnetic exchange interactions between Fe-Fe pairs, Fe-Ni pairs and Ni-Ni pairs in FeNi alloys with tetragonal bct structure (left) and random structure (right) as a function of the distance $d/a$ between the pairs of atoms $i$ and $j$ ($a$ is the lattice constant).} \label{fig:exchange_interactions}
\end{figure*}

The exchange interactions $J_{ij}$ for the ordered and random phases of FeNi for the Fe-Fe, Fe-Ni and Ni-Ni pairs are shown in Fig. \ref{fig:exchange_interactions}. The changes in the exchange interaction $J_{ij}$ and local magnetic moment play an important role for the magnetic free energy and Curie temperature. It is found that the contributions of the exchange interactions are very sensitive to the local environments of the atoms. The strongest interactions are between Fe-Fe pairs at the first nearest neighbor, while the lowest interactions are for Ni-Ni.

Panels (a) and (b) in Fig. \ref{fig:exchange_interactions} show a quite different tendencies for the Fe-Fe pairs at first nearest neighbor distance for the $L1_0$ and random phases. In the $L1_0$ phase, the positive interactions at the first coordination shell are decreased with increasing the volume, while opposite tendency is found for random FeNi. This volume dependence of exchange parameters is in agreement with Ref. \cite{Ruban2005FeNi}.

Magnetic moments tend to grow as a function of volume due to the correlation between electrons, which favors parallel spins. Increasing negative pressure typically means decreasing overlap between electronic orbitals. The volume dependency of the orbital overlap integrals will then affect the volume dependency of the $J_{ij}$ parameters. It is not easy to predict how the $J_{ij}$ parameters will change as a function of negative pressure. Ref.\cite{Wang2010exchange} found that in pure bcc Fe the pressure dependence of the $J_{ij}$ parameters is very complex. In the ordered L$1_0$ structure, each Fe atom only has four first nearest neighbour Fe atoms, which means that the total first shell Fe-Fe overlapping vanishes quickly. This explains why the first shell Fe-Fe $J_{ij}$ values decrease rapidly as a function of negative pressure. For the random phase the first shell Fe-Fe $J_{ij}$ values are seen to slightly increase, so that at the highest negative pressure the $J_{ij}$ values between the ordered and random cases are comparable in magnitude ($\approx \SI{13}{\milli\electronvolt}$). This seems reasonable, because at that pressure the magnetic moments of the ordered and random phases are equal (Fig. \ref{fig:fig3_mag}). In general, decreasing $J_{ij}$ interactions should lead to a decreasing Curie temperature. The effect of decreasing $J_{ij}$ and increasing magnetic moments partly cancel each other in the spin Hamiltonian.

\begin{figure}[ht]
	\centering
	\includegraphics[width=1.0\linewidth]{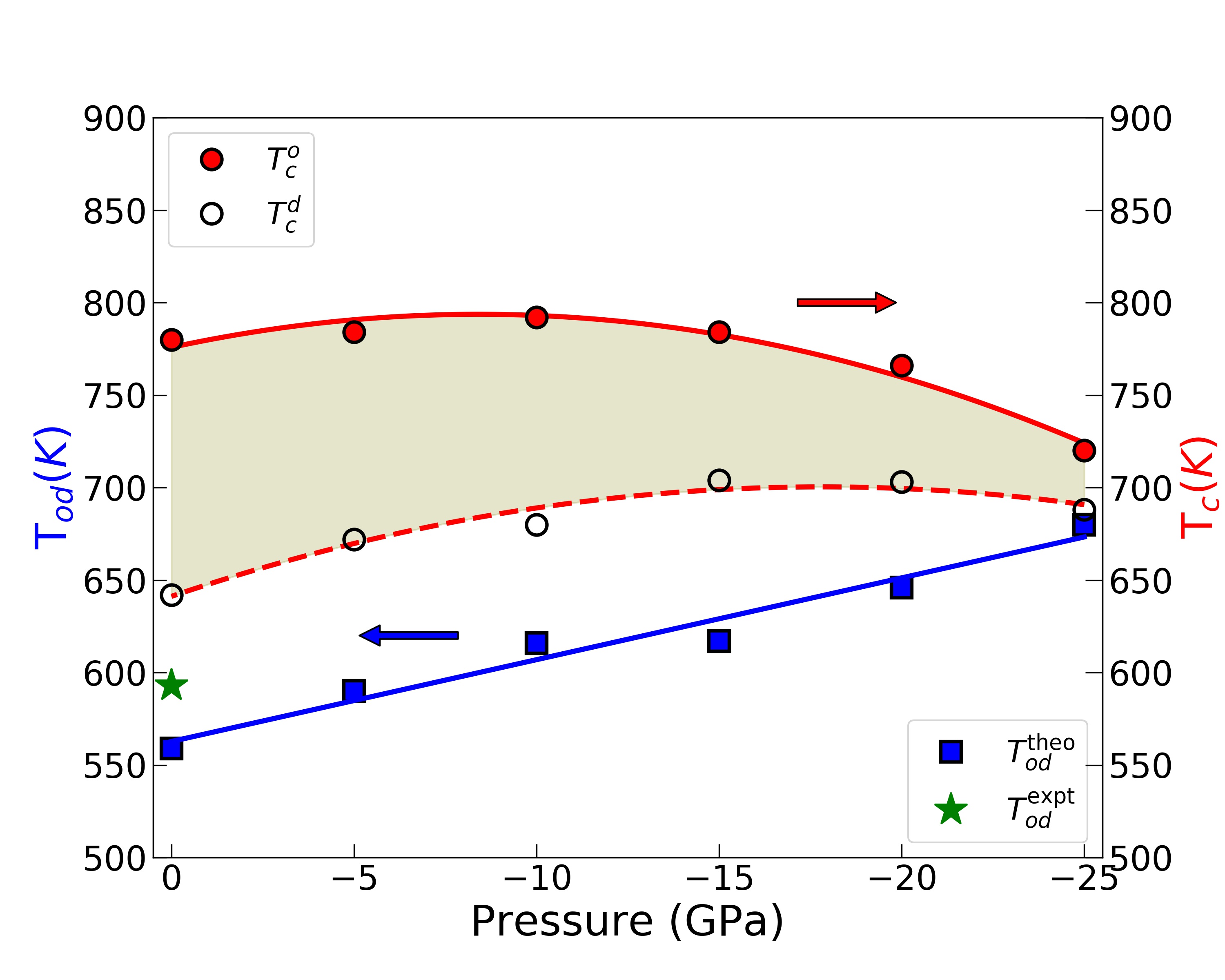}
	\caption{(Color online) Variation of the order-disorder and magnetic transition temperatures of FeNi with pressure. Open and closed circles show the Curie temperatures of ordered and disordered states, respectively, and squares show the chemical ordering temperature. Experimental ordering temperature is shown for reference \cite{Albertsen1981tetragonal}. }\label{fig:fig4_Tod}
\end{figure}

In Fig. \ref{fig:fig4_Tod}, the chemical order-disorder ($T_{od}$) and magnetic ($T_c$) transition temperatures are shown as a function of pressure. This is the central result of our investigation. Compared to the volume  at \SI{0}{\giga\pascal}, the volume of $L1_0$ FeNi at \SI{-25}{\giga\pascal} corresponds to a good approximation to the volume of FeNiN reported by Goto \emph{et al.} \cite{Goto2017}. The \emph{ab initio} theory employed here performs well in reproducing the experimental order-disorder transition temperature at \SI{0}{\giga\pascal}. Most importantly, we find that the chemical ordering temperature is increased with negative pressure. Within the present pressure interval, we find an increased value of the chemical order-disorder transition temperature of approximately \SI{121}{K}. In addition, the Curie temperatures of $L1_0$ and random FeNi phases vary smoothly with negative pressure.

\begin{figure}[ht]
	\centering
	\includegraphics[width=1.0\linewidth]{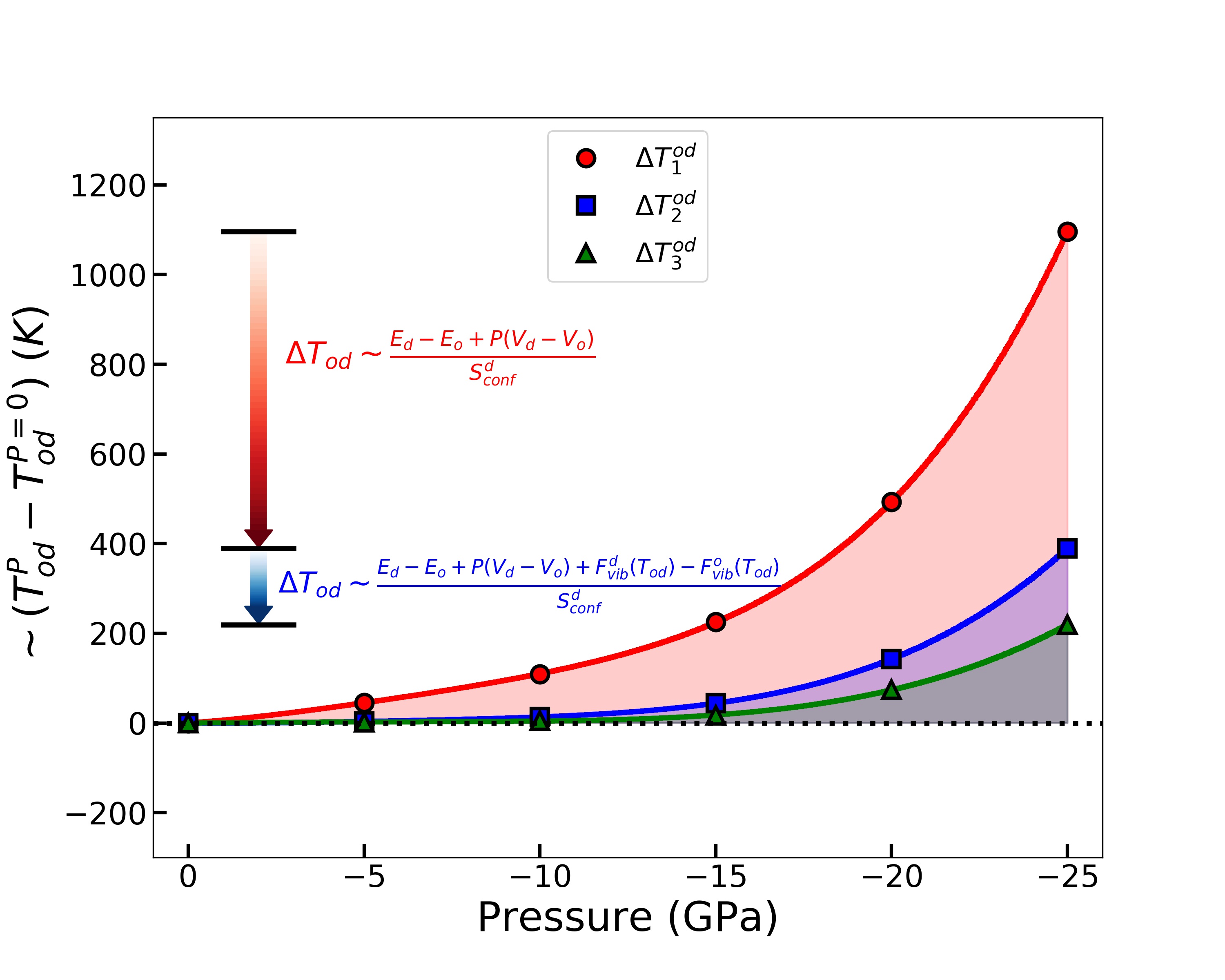}
	\caption{(Color online) The ordering temperature of FeNi alloy upon negative pressure estimated from fully ordered and disordered states. $\Delta T_{od}$ are calculated relative to the ordering temperature at zero pressure. $\Delta T^{od}_1$ considers only the configurational effects on the Gibbs energy ($E_{0K} - TS_{conf}$), $\Delta T^{od}_2$ includes the configurational and $PV$ effects ($E_{0K} - TS_{conf} + PV$), and $\Delta T^{od}_3$ includes the configuration, $PV$ and vibration effects ($E_{0K} - TS_{conf} + PV + F_{vib}$). }\label{fig:fig5_Tod} 
\end{figure}

The changes in the ordering temperature upon negative pressure can be understood by monitoring the components of the Gibbs free energy. Compared to the other effects (vibrational, electronic and magnetic effects), the configurational term ($ E_{0K} (V, \eta) - T S_{conf} (\eta) + PV$) is found to have the largest impact on the change of the ordering temperature. We notice that the configurational entropy of fully disordered state remains constant as a function of pressure and hence the change in the ($E_{0K} + PV$) term plays the key role in determining the pressure dependence of the chemical transition temperature. To illustrate this, in Fig. \ref{fig:fig5_Tod} we show the ordering temperatures relative to the zero-pressure ordering temperature. The approximate values are obtained by considering merely the configurational effects ($E_{0K} - TS_{conf}$), adding the pressure term ($E_{0K} - TS_{conf} + PV$), and adding both the pressure and vibrational effects ($E_{0K} - TS_{conf} + F_{vib} + PV$). Their contribution to ordering is different but the internal energy is found to dominate the overall pressure effect. The increase in the internal energy difference with negative pressure in turn is due to the different equations of state for the ordered and disordered phases. Namely, the thermodynamically stable ordered phase has slightly larger bulk modulus, meaning that the ordered phase reaches a particular (negative) pressure level after a smaller volume expansion compared to that of the disordered phase. The substantial internal energy contribution to the increase of $T_{od}$ is reduced by the $PV$ and vibrational terms to about \SI{219}{K} at \SI{-25}{\giga\pascal}. The sum of the electronic entropy and magnetic effects further decrease the pressure-induced transition temperature increment from 219 K to 121 K.

The increased ordering temperature in FeNi at negative pressures should promote diffusion. The diffusion near grain boundaries and dislocation-core in FeNi was estimated experimentally and for the activation energy 3.275 eV/atom ($\sim$ 316 kJ/mol) was reported \cite{Lee2014formation}. Assuming this value at all pressures, we find that the \SI{121}{K} increase of the ordering temperature results in about 3000 times larger diffusion at \SI{-25}{\giga\pascal} than at 0 pressure. For a more realistic change of the diffusion upon lattice expansion, however, one should also consider the volume effect on the activation energy, which is neglected in the above estimate. 

It is interesting to understand the relationship between Curie temperatures and chemical ordering temperatures. In the present study, the exchange parameters were used to calculate the Curie temperatures at the peaks using the Binder cumulant method \cite{Binder1981finite}. This technique has been successfully used in many studies of magnetic materials \cite{Huang2016HEA, Ruban2005FeNi, Edstrom2014}.
As seen in Fig. \ref{fig:fig4_Tod}, with expanding volume (negative pressure), the changes in $J_{ij}$ lead to a higher $T_c$ in L1$_0$ FeNi and a lower $T_c$ in random phases.
This is due to the lower magnetic ground state energy in the ordered phase compared to that in the random phase.
Since at very high temperatures the magnetic energy is zero for both phases, the temperature induced change in the magnetic energy must be greater for the ordered phase, which means that the magnetic entropy of the ordered phase must be greater.
Additionally, because the magnetic energy increases as a function of temperature, the magnetic contribution to free energy of FeNi is positive and therefore the magnetic entropy serves to increase the ordering temperature.

In conclusion, we have used first-principles alloy theory and Monte Carlo simulations to analyze the negative pressure effect on the chemical and magnetic ordering transformations of tetragonal FeNi alloys. We have demonstrated the increasing order-disorder transition in FeNi by negative pressure. Our findings contribute to the understanding of the parameter space that controls the order-disorder transformation in this technologically important material, which is significant in order to find growth conditions that stabilize the $L1_0$ phase. Monte Carlo simulations have been carried out to analyse the negative pressure effect on the chemical ordering transformation and the magnetic transition of tetragonal FeNi alloys based on the first-principles theory. 

The present results confirm a strong dependence of the transition temperature on negative pressure. The experimental order-disorder transition temperature is accurately reproduced at \SI{0}{\giga\pascal} and the chemical ordering temperature is increased linearly by approximately \SI{121}{K} with increasing the (negative) pressure from \SI{0}{\giga\pascal} to \SI{-25}{\giga\pascal}. This phenomenon is in agreement with the observations made on nitriding the FeNi by NITE technology \cite{Goto2017}. The insertion of nitrogen in FeNi, results in high degree of long-range order parameter (0.71), which is consistent with the increasing ordering temperature (\SI{680}{K}) at \SI{-25}{\giga\pascal} predicted in the present study. We conclude that the pressure-induced ordering temperature in $L1_0$ FeNi is the primary effect to understand the NITE process. By identifying this effect, we delineate a route for synthesizing ordered FeNi alloys for magnetic applications.

\section*{Acknowledgements}

%Computational resources were provided by the Swedish National Infrastructure for Computing (SNIC) at Link\"oping.
The computations were performed on resources provided by the Swedish National Infrastructure for Computing (SNIC) at Link\"oping.

\section*{Disclosure statement}

No potential conflict of interest was reported by the authors.

\section*{Funding}

The work was supported by the Swedish Research Council (VR), the Swedish Foundation for Strategic Research (SSF), the Carl Tryggers Foundations, the Swedish Innovation Agency (VINNOVA), the Hungarian Scientific Research Fund (OTKA 128229), the Swedish Energy Agency, eSSENCE and STandUPP. 

%\section{ORCID}
%Levente Vitos https://orcid.org/0000-0003-2832-3293

%\section*{Notes on contributor(s)}

%An unnumbered section, e.g.\ \verb"\section*{Notes on contributors}", may be included \emph{in the non-anonymous version} if required. A photograph may be added if requested.

%\section*{Nomenclature/Notation}

%An unnumbered section, e.g.\ \verb"\section*{Nomenclature}" (or \verb"\section*{Notation}"), may be included if required, before any Notes or References.

%\bibliographystyle{tfnlm}
%\bibliography{references}

\begin{thebibliography}{10}
	\providecommand{\url}[1]{\normalfont{#1}}
	\providecommand{\urlprefix}{Available from: }
	
	\bibitem{Kojima2014}
	Kojima~T, Ogiwara~M, Mizuguchi~M, et~al. {Fe--Ni} composition dependence of
	magnetic anisotropy in artificially fabricated {L1$_0$}-ordered {FeNi} films.
	J Phys Condens Matter. 2014;\hspace{0pt}26(6):064207.
	
	\bibitem{Skomski2013Future}
	Skomski~R, Manchanda~P, Kumar~P, et~al. Predicting the future of
	permanent-magnet materials. IEEE Trans Magn.
	2013;\hspace{0pt}49(7):3215--3220.
	
	\bibitem{Lewis2014Inspired}
	Lewis~LH, Mubarok~A, Poirier~E, et~al. Inspired by nature: investigating
	tetrataenite for permanent magnet applications. J Phys Condens Matter.
	2014;\hspace{0pt}26(6):064213.
	
	\bibitem{Lewis2014magnete}
	Lewis~LH, Pinkerton~FE, Bordeaux~N, et~al. De magnete et meteorite: cosmically
	motivated materials. IEEE Magn Lett. 2014;\hspace{0pt}5:1--4.
	
	\bibitem{Yodoshi2018effects}
	Yodoshi~N, Ookawa~S, Yamada~R, et~al. Effects of nanocrystallisation on
	saturation magnetisation of amorphous {Fe76Si9B10P5}. Mater Res Lett.
	2018;\hspace{0pt}6(1):100--105.
	
	\bibitem{Frisk2017strain}
	Frisk~A, Hase~TP, Svedlindh~P, et~al. Strain engineering for controlled growth
	of thin-film {FeNi} {L1$_0$}. J Phys D Appl Phys.
	2017;\hspace{0pt}50(8):085009.
	
	\bibitem{Kojima2014Co}
	Kojima~T, Mizuguchi~M, Koganezawa~T, et~al. Addition of {Co} to
	{L1$_0$}-ordered {FeNi} films: influences on magnetic properties and ordered
	structures. J Phys D Appl Phys. 2014;\hspace{0pt}47(42):425001.
	
	\bibitem{Shima2007film}
	Shima~T, Okamura~M, Mitani~S, et~al. Structure and magnetic properties for
	{L1$_0$}-ordered {FeNi} films prepared by alternate monatomic layer
	deposition. J Magn Magn Mater. 2007;\hspace{0pt}310(2):2213--2214.
	
	\bibitem{Kojima2014film}
	Kojima~T, Ogiwara~M, Mizuguchi~M, et~al. {Fe--Ni} composition dependence of
	magnetic anisotropy in artificially fabricated {L1$_0$}-ordered {FeNi} films.
	J Phys Condens Matter. 2014;\hspace{0pt}26(6):064207.
	
	\bibitem{Goto2017}
	Goto~S, Kura~H, Watanabe~E, et~al. Synthesis of single-phase {L1$_0$}-{FeNi}
	magnet powder by nitrogen insertion and topotactic extraction. Sci Rep.
	2017;\hspace{0pt}7(1):13216.
	
	\bibitem{Jena2014study}
	Jena~AP, Sanyal~B, Mookerjee~A. Study of the effect of magnetic ordering on
	order--disorder transitions in binary alloys. J Magn Magn Mater.
	2014;\hspace{0pt}360:15--20.
	
	\bibitem{Tian2019FeNi}
	Tian~LY, Lev{\"a}m{\"a}ki~H, Eriksson~O, et~al. Density functional theory
	description of the order-disorder transformation in {Fe-Ni}. Sci Rep.
	2019;\hspace{0pt}9(1):8172.
	
	\bibitem{Vitos2000a}
	Vitos~L, Skriver~HL, Johansson~B, et~al. Application of the exact muffin-tin
	orbitals theory: the spherical cell approximation. Comput Mater Sci.
	2000;\hspace{0pt}18(1):24--38.
	
	\bibitem{Vitos2001c}
	Vitos~L, Abrikosov~I, Johansson~B. Anisotropic lattice distortions in random
	alloys from first-principles theory. Phys Rev Lett.
	2001;\hspace{0pt}87(15):156401.
	
	\bibitem{Vitos2007book}
	Vitos~L. Computational quantum mechanics for materials engineers: the {EMTO}
	method and applications. Springer Science \& Business Media; 2007.
	
	\bibitem{Skubic2008MC}
	Skubic~B, Hellsvik~J, Nordstr{\"o}m~L, et~al. A method for atomistic spin
	dynamics simulations: implementation and examples. J Phys Condens Matter.
	2008;\hspace{0pt}20(31):315203.
	
	\bibitem{Soven1967coherent}
	Soven~P. Coherent-potential model of substitutional disordered alloys. Phys
	Rev. 1967;\hspace{0pt}156(3):809.
	
	\bibitem{Gyorffy1972coherent}
	Gyorffy~B. Coherent-potential approximation for a
	nonoverlapping-muffin-tin-potential model of random substitutional alloys.
	Phys Rev B. 1972;\hspace{0pt}5(6):2382.
	
	\bibitem{Murnaghan1944}
	Murnaghan~FD. {The Compressibility of Media under Extreme Pressures}.
	Proceedings of the National Academy of Sciences of the United States of
	America. 1944;\hspace{0pt}30(9):244--7.
	
	\bibitem{Birch1947}
	Birch~F. {Finite elastic strain of cubic crystals}. Phys Rev.
	1947;\hspace{0pt}71(11):809--824.
	
	\bibitem{Ruban2005FeNi}
	Ruban~AV, Katsnelson~M, Olovsson~W, et~al. Origin of magnetic frustrations in
	{Fe-Ni} {Invar} alloys. Phys Rev B. 2005;\hspace{0pt}71(5):054402.
	
	\bibitem{Wang2010exchange}
	Wang~H, Ma~PW, Woo~C. Exchange interaction function for spin-lattice coupling
	in bcc iron. Phys Rev B. 2010;\hspace{0pt}82(14):144304.
	
	\bibitem{Albertsen1981tetragonal}
	Albertsen~J. Tetragonal lattice of tetrataenite (ordered fe-ni, 50-50) from 4
	meteorites. Phys Scripta. 1981;\hspace{0pt}23(3):301.
	
	\bibitem{Lee2014formation}
	Lee~S, Edalati~K, Iwaoka~H, et~al. Formation of {FeNi} with {L$1_0$}-ordered
	structure using high-pressure torsion. Philos Mag Lett.
	2014;\hspace{0pt}94(10):639--646.
	
	\bibitem{Binder1981finite}
	Binder~K. Finite size scaling analysis of {Ising} model block distribution
	functions. Z Phys, B, Condens Matter. 1981;\hspace{0pt}43(2):119--140.
	
	\bibitem{Huang2016HEA}
	Huang~S, Li~W, Li~X, et~al. Mechanism of magnetic transition in
	{FeCrCoNi}-based high entropy alloys. Mater Des. 2016;\hspace{0pt}103:71--74.
	
	\bibitem{Edstrom2014}
	Edstr{\"o}m~A, Chico~J, Jakobsson~A, et~al. Electronic structure and magnetic
	properties of {L1$_0$} binary alloys. Phys Rev B.
	2014;\hspace{0pt}90(1):014402.
	
\end{thebibliography}

%\subsubsection*{Counts of words} 
%\wordcount

\end{document}